\documentclass[11pt]{article}
\usepackage{amsmath,amsfonts}

\begin{document}

\title{\textbf{Effective gravity from a quantum gauge theory in Euclidean
space-time}}
\author{\textbf{R.~F.~Sobreiro}\thanks{%
sobreiro@cbpf.br} \ , \textbf{V.~J.~Vasquez Otoya}\thanks{%
vjose@cbpf.br}\\
%EndAName
\\
\textit{CBPF, Centro Brasileiro de Pesquisas F\'{\i}sicas,} \and
\textit{Rua Xavier Sigaud 150, 22290-180 Urca,} \textit{Rio de
Janeiro, Brasil} }
\date{}
\maketitle

\begin{abstract}
\noindent We consider a $SO(d)$ gauge theory in an Euclidean
$d$-dimensional space-time, which is known to be renormalizable to
all orders in perturbation theory for $2\le{d}\le4$. Then, with the
help of a space-time representation of the gauge group, the gauge
theory is mapped into a curved space-time with linear connection.
Further, in that mapping the gauge field plays the role of the
linear connection of the curved space-time and an effective metric
tensor arises naturally from the mapping. The obtained action, being
quadratic in the Riemann-Christoffel tensor, at a first sight,
spoils a gravity interpretation of the model. Thus, we provide a
sketch of a mechanism that breaks the $SO(d)$ color invariance and
generates the Einstein-Hilbert term, as well as a cosmological
constant term, allowing an interpretation of the model as a modified
gravity in the Palatini formalism. In that sense, gravity can be
visualized as an effective classical theory, originated from a well
defined quantum gauge theory. We also show that, in the four
dimensional case, two possibilities for particular solutions of the
field equations are the de Sitter and Anti de Sitter space-times.
\end{abstract}

\section{Introduction}

Since the beginning of the last century many efforts to give gravity
a quantum description have been developed. However, a complete
consistent quantum theory of gravity still lacks. In particular,
attempts to quantize general relativity have failed since it is a
nonrenormalizable theory where higher order derivative counterterms
already appear at one-loop quantum corrections
\cite{'tHooft:1974bx,Deser:1974cz,Deser:1974cy}. In order to solve
this problem gravity has been described by several alternative
approaches. Some of them are: The higher derivative gravities
\cite{Stelle:1976gc}; The loop quantum gravity (LQG)
\cite{Rovelli:2004tv,Smolin:2004sx}; The Einstein-Cartan gauge
formalism \cite{Utiyama:1956sy,Kibble:1961ba}; The interpretation
that gravity would be an effective theory originated from more
general theories such as superstring theories
\cite{Ferrara:1986qw,Deser:1986ww}. It is important to highlight
though that it is not our intention to give a complete description
on those issues, for that we refer to the cited bibliography and
references therein.\\\\ To discuss quantum aspects of gravity, let
us take under consideration the pure quantum mechanical point of
view. It is reasonably unanimous that Quantum Physics is the
fundamental theory while Classical Physics is a particular case of
the quantum theory. From this point of view, quantization methods
employed to achieve a quantum model from a classical model are just
\emph{ans\"atze} or algorithms that might not always work. A
renowned example is a general dissipative system. This is the main
quantum mechanics point, implying that the word \emph{quantize}
should not be taken as a fundamental principle
\cite{Gitman:1990qh,Bolivar}. In quantum field theory, the picture
is not different. The quantization methods are just \emph{ans\"atze}
successfully utilized to obtain a quantum theory from a classical
model. The starting point would be the fundamental quantum theory
from where, in some suitable limits, one recovers the classical
equivalent theory. This might be the reason why gravity resists to
be quantized by the usual quantization methods, requiring then a
more fundamental understanding of quantum theory.\\\\ Thus, based on
the fundamental character of the quantum theory and on the success
of gauge theories when utilized to describe the fundamental
interactions, we opt, in this work, for the idea that gravity can be
visualized as an effective theory originated from a simple,
non-Abelian, four-dimensional Euclidean quantum gauge field theory.
To be more specific, we start our investigation on the proposed idea
by considering a $SO(d)$ gauge theory in an Euclidean
$d$-dimensional space-time\footnote{Here we are considering a
$d$-dimensional space just for generality purposes, the main focus
being the case $d=4$.}. Under this prescription the graviton is
simply defined as the gauge field, $A^a_i$. We recall that an
essential ingredient of this description is that there is no curved
space in the quantum sector of the theory. In fact, the Euclidean
space-time is a motivation to define the path integral allowing a
useful consistent scenario to perform quantum computations.\\\\ Once
we have established the properties of the theory at quantum level,
we can relate it with a curved space-time through a mapping from the
Euclidean gauge theory to a dynamical space-time with
$GL(d,\mathbb{R})$ symmetry. For that, one can use a representation
of the $SO(d)$ gauge group borrowing the Euclidean space-time
indices, then the group index $a$ can be represented by a pair of
antisymmetric space-time indices,
$A_i^{\phantom{i}jk}=-A_i^{\phantom{i}kj}$. In this mapping the
Euclidean space is identified with a curved space through
transformation matrices $e_i^\mu$, which are related to an effective
metric tensor $g$. The gauge field is mapped into a linear
connection of the space-time
$A_{ij}^{\phantom{ij}k}\mapsto\Gamma_{\mu\nu}^{\phantom{\mu\nu}\alpha}$
which is independent of the metric tensor. Thus, the Euclidean gauge
theory is mapped into a dynamical space-time in the Palatini
formalism. The original action, $\int\textbf{F}^2$, is mapped into
an action quadratic in the Riemann tensor
$\int\sqrt{g}\;\textbf{R}\cdot \textbf{R}$.\\\\ In order to relate
the dynamical curved space with gravity, it would be useful to
generate the Einstein-Hilbert (EH) term in the action. However, the
term $\textbf{R}\cdot\textbf{R}$ still carries the global gauge
symmetry of the original gauge theory while the EH term,
$R=R_{\mu\nu}^{\phantom{\mu\nu}\mu\nu}=\mathrm{Tr}\;\textbf{R}$,
breaks color invariance. Thus, we provide a sketch of a color
symmetry breaking mechanism which generates the EH term. The first
feature of the color symmetry breaking is the appearance of physical
spin-2 excitations originated from the antisymmetry structure of the
color indices. Before the breaking, those excitations were hidden
behind the color symmetry and thus were unphysical. It is important
to know that this mechanism is employed in the quantum sector of the
gauge theory, \emph{i.e}, the mapping is only performed after the
mechanism is worked out. This detail ensures the existence of a
framework for well defined quantum computations, as an Euclidean QFT
is. An extra feature of the mechanism is that a cosmological
constant (CC) appears naturally in the resulting action. We remark
at this point, and will be pointed further, that this mechanism is
merely a possible sketch. In fact, further investigation on the
formalization of the present mechanism (or
other) is needed and are currently under investigation \cite{work}.\\\\
Finally, once the previously mechanism is worked out, the mapping
can be performed. The obtained action for the dynamical space-time
is now constituted by a quadratic term in the Riemann tensor, the EH
term and the CC term. This new action can now be related with
gravity, in the Palatini formalism. In the particular case of four
dimensions, the equations of motion provide the de Sitter space-time
($dS_4$) or the Anti de Sitter ($AdS_4$) as the fundamental
space-time.\\\\ This work is organized as follows: In section 2, we
will show how a $SO(d)$ gauge model, in $d$-dimensional Euclidean
space-time, can be mapped into a dynamical curved space. The
resulting dynamical space-time enjoys $GL(d,\mathbb{R})$ symmetry
and $SO(d)$ gauge symmetry (color invariance). In section 3, we will
argue that, to obtain an effective gravity theory from the previous
method, a color symmetry breaking mechanism is required. In section
4, a color symmetry breaking mechanism will be sketched and, due to
this mechanism, the EH term is generated together with the CC. Still
in this section, we will discuss the resulting classical field
equations. Finally, the conclusions will be displayed in section 5.

\section{From $SO(d)$ gauge theory to dynamical curved space-time}

We start this section with a small review of the definitions,
notations and properties of the $SO(d)$ gauge theory.

\subsection{Generalities of the $SO(d)$ gauge theory}

Let us consider the following principal bundle
\begin{equation}
\mathcal{P}\equiv\{SO(d)\;;\;\mathbb{R}^d\;;\;\mathcal{A}\}\;,\label{pb1}
\end{equation}
where $SO(d)$ is the group of orthogonal matrices characterizing the
fiber and the structure group of the fiber. There are $D=d(d-1)/2$
generators of this group which we label by $\lambda^a$,
$a\in\{1,2,\ldots,D\}$. The algebra is a typical Lie algebra,
\begin{equation}
[\lambda^a,\lambda^b]=f^{abc}\lambda^c\;,\label{alg0}
\end{equation}
where $f^{abc}$ represents the structure constants of the group.
Also, we denote the elements of the group by
$u=e^{\omega^a\lambda^a}$. The total space of the principal bundle
(\ref{pb1}) is chosen to be an Euclidean $d$-dimensional space,
$\mathbb{R}^d$, which enjoys a global $O(d)$ symmetry. The set
$\mathcal{A}$ is the space of the algebra-valued connections, the
base space\footnote{The first part of the Latin alphabet,
$\{a,b,c,\ldots,h\}$, labels the group indices while the second
part, $\{i,j,k,\ldots,z\}$, will be used to label the space-time
coordinates.},
\begin{equation}
A_i=A_i^a\lambda^a\;.\label{conn1}
\end{equation}
The homeomorphisms of the principal bundle are characterized by the
gauge transformations
\begin{equation}
A_i\rightarrow{A}_i+u^\dagger{D}_iu\;,\label{gauge1}
\end{equation}
where the covariant derivative is just
\begin{equation}
D_i\cdot=\partial_i\cdot+[{A}_i,\cdot]\;.\label{covder1}
\end{equation}
Infinitesimal gauge variations are then defined as
\begin{equation}
\delta{A}_i=D_i\omega\;.\label{gauge2}
\end{equation}
The gauge invariant action is
\begin{equation}
S_{\mathrm{YM}}=\frac{1}{4\kappa^2}\mathrm{Tr}\int{d^dx}F_{ij}F^{ij}\;,\label{action1}
\end{equation}
where the field strength (fiber curvature) is given by
\begin{equation}
F_{ij}=F_{ij}^a\lambda^a=\partial_iA_j-\partial_jA_i+[A_i,A_j]\;.\label{curv0}
\end{equation}
The quantity $\kappa$ is the coupling constant.\\\\ To quantize the
model a gauge fixing is usually needed. Here we choose the simplest
covariant gauge, the Landau gauge $\partial^iA_i=0$. With that
purpose we add to the gauge invariant action (\ref{action1}) a gauge
fixing term
\begin{equation}
S=S_{\mathrm{YM}}+S_{\mathrm{gf}}\;,\label{action2}
\end{equation}
where
\begin{equation}
S_{\mathrm{gf}}=\mathrm{Tr}\int{d^dx}\;\left(b\;\partial^iA_i+\bar{c}\;\partial^iD_ic\right)\;.\label{action3}
\end{equation}
The fields $c=c^a\lambda^a$ and $\bar{c}=\bar{c}^a\lambda^a$ are,
respectively, the Faddeev-Popov ghost and antighost fields while
$b=b^a\lambda^a$ is the Lautrup-Nakanishi field which plays the role
of a Lagrange multiplier enforcing the Landau gauge condition. The
action (\ref{action2}) enjoys renormalizability, \emph{i.e.}, the
model is consistent at quantum level. In fact, as discussed in
Ap.\ref{ap1}, the renormalizability of action (\ref{action2}) can be
established to all orders in
perturbation theory for $2\le d\le4$.\\\\
After gauge fixing, a global $SO(d)$ gauge symmetry survives, the so
called color invariance. This invariance is characterized by the
non-observational character of the group indices. In terms of a Ward
identity, color invariance is described by
\begin{equation}
\int{d^dx}\left(\left[A_i,\frac{\delta
S}{\delta{A}_i}\right]+\left[c,\frac{\delta
S}{\delta{c}}\right]+\left[\bar{c}, \frac{\delta
S}{\delta\bar{c}}\right]+\left[b,\frac{\delta
S}{\delta{b}}\right]\right)=0\;.\label{color1}
\end{equation}

\subsection{Space-time representation}

Unless the contrary is said, we will not consider the gauge fixing
term for the rest of the article. Thus, from now on, we are dealing
with the gauge invariant action (\ref{action1}) and global color
invariance is now described by
\begin{equation}
\int{d^dx}\left[A_i,\frac{\delta
S_{\mathrm{YM}}}{\delta{A}_i}\right]=0\;.\label{color2}
\end{equation}\\\\ First, we will explore the fact that the $SO(d)$ group, in an
Euclidean space-time, can borrow the space-time structure to fix a
representation of the group. For that, we observe that the dimension
of the total space $\mathbb{R}^d$ coincides with the Casimir $d$ of
the group $SO(d)$, which means that the generators can be
represented as a set of $(d\times d)$-matrices. Also, the group
dimension, $D$, coincides with the number of independent elements of
an antisymmetric $(d\times d)$-matrix. Those properties allow a
unique representation of the gauge group where the generators are in
the form of matrices labeled by space-time indices,
\begin{eqnarray}
\lambda^a&\mapsto&\lambda^{ij}\;,\nonumber\\
f^{abc}&\mapsto&f^{ijklmn}\;,\label{map1}
\end{eqnarray}
where
\begin{eqnarray}
\lambda^{ij}&=&-\lambda^{ji}\;,\nonumber\\
f^{ijklmn}&=&-\frac{1}{2}\left[\left(\delta^{il}\delta^{mj}-
\delta^{jl}\delta^{mi}\right)\delta^{kn}+\left(\delta^{jk}\delta^{im}-
\delta^{ik}\delta^{mj}\right)\delta^{ln}\right]\;.\label{map2}
\end{eqnarray}
Thus
\begin{eqnarray}
A_i^a\lambda^a&=&A_{ij}^{\phantom{ij}k}\lambda^j_{\phantom{j}k}\;,\nonumber\\
F_{ij}^a\lambda^a&=&F_{ijk}^{\phantom{ijk}l}\lambda^k_{\phantom{k}l}\;.
\label{map3}
\end{eqnarray}\\\\ In the above described representation the action (\ref{action1})
reads
\begin{equation}
S_{\mathrm{YM}}=\frac{1}{4\kappa^2}\int{d^dx}F_{ijk}^{\phantom{ijk}l}F^{ijk}_{\phantom{ijk}l}\;,\label{action4}
\end{equation}
where
\begin{equation}
F_{ijk}^{\phantom{ijk}l}=\partial_iA_{jk}^{\phantom{jk}l}-
\partial_jA_{ik}^{\phantom{ik}l}+A_{im}^{\phantom{im}l}A_{jk}^{\phantom{jk}m}-A_{jm}^{\phantom{jm}l}A_{ik}^{\phantom{ik}m}\;.\label{fs1}
\end{equation}
Thus, we are dealing with a gauge theory, in an Euclidean space-time
with a representation where the color indices are represented by
space-time indices.\\\\ As previously pointed, the color invariance
has a fundamental role in this article, therefore, for further use,
let us write the color symmetry Ward identity in the space-time
representation
\begin{equation}
\int{d^dx}\left(A_{ki}^{\phantom{ki}n}\delta_{mj}-A_{kj}^{\phantom{kj}n}\delta_{mi}\right)\frac{\delta
S_{\mathrm{YM}}}{\delta{A}_{km}^{\phantom{km}n}}=0\;.\label{color2}
\end{equation}

\subsection{Pseudo-vielbein and curved space-time}

Looking at the field strength expression in (\ref{fs1}), one can
recognize an extreme similarity with a Riemann-Christoffel tensor of
a curved space-time. Notwithstanding, one can think in absorb the
gauge field as a geometrical structure of the space-time and end up
in an effective geometrical theory, completely equivalent to a gauge
theory. In fact, we can perform a mapping from the Euclidean space
to a curved space\footnote{The Greek indices label the coordinates
of the curved space time.},
\begin{equation}
\{x_i\}\mapsto\{x_\mu,x^\mu\}\;,\label{map4}
\end{equation}
by introducing the pseudo-vielbein and its inverse
\begin{eqnarray}
e^i_\mu&=&\frac{\partial{x}^i}{\partial{x}^\mu}\;,\nonumber\\
\bar{e}_i^\mu&=&\frac{\partial{x}^\mu}{\partial{x}^i}\;.\label{pv1}
\end{eqnarray}
Thus
\begin{eqnarray}
dx^i&=&e^i_\mu dx^\mu\;,\nonumber\\
dx^\mu&=&\bar{e}_i^\mu dx^i\;.\label{map5}
\end{eqnarray}
The name \emph{pseudo-vielbein}\footnote{For economical purposes we
will call the \emph{pseudo-vielbein} simply by \emph{p-vielbein}.}
is here employed due to the fact that $e$ is not the vielbein of the
EC theory of gravity \cite{Utiyama:1956sy,Kibble:1961ba}, as it will
become evident in what follows. In that sense, the flat space is
completely independent of the curved space. Differently of the EC
formalism the p-vielbein does not define a reference frame which
travels tangently on the curved space. In here, the p-vielbein is a
point-to-point identification of the two spaces which have no
direct geometrical relation. We will return to this issue soon.\\\\
The requirement of metric preservation in both spaces implies on the
existence of a symmetric metric tensor $\{g_{\mu\nu},g^{\mu\nu}\}$,
in fact the following relations hold,
\begin{eqnarray}
e_\mu^ie_{i\nu}&=&g_{\mu\nu}\;,\nonumber\\
\bar{e}_i^\mu\bar{e}_\mu^j&=&\delta_i^j\;,\nonumber\\
e_i^\mu\bar{e}_\mu^j&=&\delta_i^j\;,\nonumber\\
e_i^\mu\bar{e}_\nu^i&=&\delta_\nu^\mu\;,\label{g0}
\end{eqnarray}
where the first two are related to the metric invariance in both
spaces. In that sense the p-vielbein is equivalent to the usual
vielbein, which justifies the name. However, here, $e$ is
essentially a mapping field.\\\\ For the ordinary derivative and for
the gauge field the mapping occurs according to
\begin{eqnarray}
\partial_i&=&\bar{e}^\mu_i\partial_\mu\;,\nonumber\\
A_{ij}^{\phantom{ij}k}&=&\bar{e}_i^\mu\bar{e}_j^\nu
e^k_\alpha\Gamma_{\mu\nu}^{\phantom{\mu\nu}\alpha}-\bar{e}^\mu_ie_\nu^k\partial_\mu\bar{e}_j^\nu\;.\label{map7}
\end{eqnarray}
The quantity $\Gamma$ is identified with the connection of the
curved space.\\\\
Now we return to the geometrical properties of the p-vielbein. In
the EC formalism the vielbein connects the curved space with a
tangent space. This association implies that the partial derivatives
do not commute in flat space while the derivatives in curved space
do commute. In our formalism, the original flat space-time is not a
tangent space of the curved space. They are essentially independent
spaces, modulo the mapping through $e_i^\mu$. Thus, the partial
derivative commutes in Euclidean space as well as in curved space,
\begin{eqnarray}
\left[\partial_i,\partial_j\right]&=&0\;,\nonumber\\
\left[\partial_\mu,\partial_\nu\right]&=&0\;.\label{com1}
\end{eqnarray}
Those commutation rules imply that the non-holonomicity coefficients
of both spaces vanish. Consequently, the p-vielbein behaves exactly
as a coordinate matrix transformation.\\\\ Finally, we wish to let
clear that there is no \emph{a priori} proof of the existence of
consistent (invertible) transformations $e^\mu_i$. For now, we
naively assume that $e_i^\mu$ exists and is invertible.
Nevertheless, this assumption will be confirmed \emph{a posteriori}
when we solve the field equations in Sec.\ref{grav}. Essentially, a
kind of proof of the existence of an invertible $e_i^\mu$ would rely
on solving the field equations in the curved space-time and find an
invertible metric tensor as a consistent solution.

\subsection{Action for the dynamical curved space}

A problem to be faced at this point is that we must generate a
scalar action on the curved space, which means that $F$, in the
curved space, should be a tensor. Substituting (\ref{map7}) in
(\ref{fs1}) and making use of the relations (\ref{com1}) we find
\begin{equation}
F_{ijk}^{\phantom{ijk}l}=\bar{e}_i^\mu\bar{e}_j^\nu \bar{e}_k^\alpha
e^l_\beta
R_{\mu\nu\alpha}^{\phantom{\mu\nu\alpha}\beta}\;,\label{map9}
\end{equation}
where $R_{\mu\nu\alpha}^{\phantom{\mu\nu\alpha}\beta}$ is a four
rank tensor in the form
\begin{equation}
R_{\mu\nu\alpha}^{\phantom{\mu\nu\alpha}\beta}=\partial_\mu\Gamma_{\nu\alpha}^{\phantom{\nu\alpha}\beta}-
\partial_\nu\Gamma_{\mu\alpha}^{\phantom{\mu\alpha}\beta}+
\Gamma_{\mu\gamma}^{\phantom{\mu\gamma}\beta}\Gamma_{\nu\alpha}^{\phantom{\nu\alpha}\gamma}-
\Gamma_{\nu\gamma}^{\phantom{\nu\gamma}\beta}\Gamma_{\mu\alpha}^{\phantom{\mu\alpha}\gamma}\;,\label{R}
\end{equation}
recognized as the Riemann-Christoffel tensor. At the action
(\ref{action4}), this transformation results in
\begin{equation}
S_{YM}^{\mathrm{e}}=\frac{1}{4\kappa^2}\int{d^dx}\;\mathrm{e}\;R_{\mu\nu\alpha}^{\phantom{\mu\nu\alpha}\beta}
R^{\mu\nu\alpha}_{\phantom{\mu\nu\alpha}\beta}\;,\label{action6b}
\end{equation}
where $e=\det{e}^i_\mu=\sqrt{-\det g_{\mu\nu}}$ comes from the
Jacobian of the transformation (\ref{map5}). The action
(\ref{action6b}) displays color invariance under the $SO(d)$ group
characterized by the following functional identity
\begin{equation}
\int{d^dx}\left(\Gamma_{\mu\nu}^{\phantom{\mu\nu}\alpha}\delta_{\beta\gamma}-
\Gamma_{\mu\gamma}^{\phantom{\mu\gamma}\alpha}\delta_{\beta\nu}\right)\frac{\delta
S_{\mathrm{YM}}}{\delta\Gamma_{\mu\beta}^{\phantom{\mu\beta}\alpha}}=0\;.\label{color3}
\end{equation}
Further, it is clear that (\ref{action6b}) enjoys $GL(d,\mathbb{R})$
symmetry, not at the gauge sector, but in the curved space-time
itself. What does it mean? It means that we started from a
renormalizable gauge theory. Due to the possibility of the
space-time representation the classical theory is then mapped into a
curved space with linear connection. The gauge field turns into the
linear connection and an effective metric tensor arises from the
p-vielbein.

\subsection{On-shell field equations}

The action (\ref{action6b}) describes the dynamics of a curved
space-time. The fundamental field being the connection $\Gamma$. The
explicit form of $\Gamma$, which dictates the class of geometry we
are dealing with, is unknown till now. Thus, at classical level, the
field equations of the connection are just
\begin{equation}
-\mathcal{D}_\mu{R}^{\mu\nu\alpha}_{\phantom{\mu\nu\alpha}\beta}-2\left[\frac{d}{2}Q_\mu+T_{\mu\kappa}^{\phantom{\mu\nu}\kappa}\right]
R^{\mu\nu\alpha}_{\phantom{\mu\nu\alpha}\beta}+
T_{\mu\kappa}^{\phantom{\mu\kappa}\nu}R^{\mu\kappa\alpha}_{\phantom{\mu\kappa\alpha}\beta}=0\;,\label{fe1}
\end{equation}
where
\begin{equation}
Q_\mu=\frac{1}{d}g^{\alpha\beta}\mathcal{D}_\mu
g_{\alpha\beta}\;,\label{weyl1}
\end{equation}
is the Weyl convector and
\begin{equation}
T_{\mu\nu}^{\phantom{\mu\nu}\alpha}=\frac{1}{2}\left(\Gamma_{\mu\nu}^{\phantom{\mu\nu}\alpha}-
\Gamma_{\nu\mu}^{\phantom{\mu\nu}\alpha}\right)\;,\label{tor1}
\end{equation}
is the torsion tensor. The covariant derivative $\mathcal{D}$ is the
usual covariant derivative of a general curved space-time with
linear connection $\Gamma$, which is different of the definition of
the covariant derivative (\ref{covder1}).\\\\ Evidently, the
p-vielbein constitutes a field and it will have its respective field
equation. However, it is important to keep in mind that the
fundamental field here is the connection which has well defined
quantum properties. The p-vielbein is a pure classical field. The
classical field equations of the connection are the usual classical
limit selecting the main contribution to the path integral. On the
other hand, the field equations for the p-vielbein are interpreted
as a minimization principle related to the vacuum energy stability.
In fact, there are several possible solutions to $e$. Nevertheless,
through the field equations, we select the p-vielbein which makes
the vacuum energy stable. In principle, it is just a matter of
minimal principles. Thus, the variation of the action
(\ref{action6b}) with respect to the metric tensor
provides\footnote{The same equation can be achieved by varying the
action (\ref{action6b}) with respect to the p-vielbein.}
\begin{equation}
\frac{1}{2}g^{\mu\nu}R^{\sigma\rho\alpha\beta}R_{\sigma\rho\alpha\beta}-
R^\mu_{\phantom{\mu}\alpha\beta\gamma}R^{\nu\alpha\beta\gamma}-
R^{\phantom{\alpha\beta\gamma}\mu}_{\alpha\beta\gamma}R_{\phantom{\nu\alpha\beta\gamma}}^{\alpha\beta\gamma\nu}=0\;.\label{fe2}
\end{equation}
Now, contracting the equation for the p-vielbein with $g_{\mu\nu}$
one finds
\begin{equation}
\frac{(d-4)}{2}R^{\sigma\rho\alpha\beta}R_{\sigma\rho\alpha\beta}=0\;.\label{fe3}
\end{equation}\\\\ We recall that the geometry involved is
determined by the field equations (\ref{fe1}) and (\ref{fe2}). The
solution of those equations will provide the geometrical properties
of the curved space. \emph{The curved space is simply a space with
single linear connection}, which generalizes several geometries such
as the Riemannian geometry and the Einstein-Cartan geometry. From
that point of view, we are dealing with the Palatini formalism, the
so called metric-affine formulation. The difference is that action
(\ref{action6b}) no longer describes gravity. In the next section,
we start the discussion whether the presented formalism can be
related to a gravity theory or not.\\\\ Exploiting, the field
equations (\ref{fe1}), (\ref{fe2}) and (\ref{fe3}) for $d\ne4$, we
observe that equation (\ref{fe3}) implies that
$R^{\sigma\rho\alpha\beta}=0$, which is also a solution of
(\ref{fe1}) and (\ref{fe2}). In other words, the case $d\ne4$ is
trivial when matter is not being considered.\\\\ For the case $d=4$
equation (\ref{fe3}) is satisfied for any non-vanishing curvature.
In that case, the nontrivial solutions of (\ref{fe1}) are allowed
and nontrivial curved spaces show up. Obviously, this phenomenon
occurs due to the nonlinear character of the action
(\ref{action6b}). Physically, this character is associated to the
self interaction of $\Gamma$.

\section{Intermediate discussion}

We wish to relate the above described theory with gravity through
the following point of view:\\\\ \emph{Quantum gravity would be
described by the action (\ref{action1}) in Euclidean space-time,
which is unitary and renormalizable. In that hypothesis, the gauge
field (\ref{conn1}) is interpreted as the graviton. Thus, at very
high energies gravity is simply a renormalizable quantum gauge
theory in Euclidean space-time, where a spin-1 gauge field plays the
role of the graviton.\\\\ On the other hand, at low energies, the
theory can be described by a deformed space-time where the gauge
field is visualized as a linear connection. In that regime an
effective metric tensor arises due to the presence of the
p-vielbein. Thus, action (\ref{action6b}) can be regarded as a kind
of effective theory of gravity. In that sense, classical gravity is
no longer a fundamental theory. The deformation of the space-time
occurs from quantum effects of a gauge theory as well as the
covariance principle of general relativity.}\\\\ We remark that the
present approach to treat gravity is essentially different of the
Einstein-Cartan formalism \cite{Utiyama:1956sy,Kibble:1961ba}. In
the EC formalism there is an initial assumption of the existence of
a curved space as the fundamental space. The tangent flat space is
used in order to define the spin connection which allows the
introduction of fermion fields with the help of the vielbein.
However, the formalism still lacks renormalizability once the EH
term is taken as the fundamental action for gravity. In our
prescription the fundamental theory is a consistent Euclidean
quantum gauge theory of spin-1 fields which, for a certain limit, is
equivalent to a pure dynamical curved space-time where no reference
to a tangent space neither spin connection is made. Thus, the curved
space is a direct consequence of the dynamics of the gauge theory
and not the contrary. Furthermore, the original flat space is not a
tangent space of the curved space, as it is evident from
(\ref{com1}).\\\\ This idea turns on essentially three issues to be
faced. First of all: \emph{Where is the EH term $\propto
R=R^{\mu\nu}_{\phantom{\mu\nu}\nu\mu}$, which ensures the relation
of a theory with gravity?} It is easy to see that this question hits
directly the color symmetry of action (\ref{action6b}). In fact, the
EH term would require contraction between group
indices and space-time indices, which breaks color invariance.\\\\
The second point concerns the spin of the physical excitations. At
quantum level, the physical excitations carries spin-1, as any gauge
theory of vector connection. Curiously, in the space-time
representation, spin-2 excitations also show up, due to the
identification of the group indices with the space-time indices.
However, due to color symmetry, those are nonphysical excitations,
which will never be observed. Thus, the question of the spin-2
physical excitations persists: \emph{Would color invariance forbid
the existence of spin-2 excitations at the physical sector of the
theory?}\\\\ Finally, the equivalence between actions
(\ref{action1}) and (\ref{action6b}), for now, due to color
invariance, seems to be just a point of view, \emph{i.e.}, is just
another way to look at the nature. It happens because color
invariance forbids the observational character of color indices. For
example, in $R_{\mu\nu\alpha\beta}$ the last two indices are related
to color invariance. Then, if one wishes to measure, directly, the
curvature of the space time, it would never been successful, since
the last two indices are non-observable. Therefore, the third
question is: \emph{Which physical effect pushes the theory to the
action (\ref{action6b}), characterizing the low energy regime of
(\ref{action1}) in terms of an observable curved space-time?}\\\\
Those questions are then closely related to the color symmetry of
action (\ref{action6b}). Further, a color symmetry breaking
mechanism seems to be necessary if one wishes to describe gravity
with the above program. With that purpose, in the next section, we
will provide a sketch of a possible color symmetry breaking
mechanism. We will show that this mechanism can generate the
physical spin-2 excitations as well as the EH term. Further, the
same mechanism would dictate the vacuum of the theory, allowing a
physical motivation for the mappings (\ref{map5}) and (\ref{map7}).\\\\
We would like to recall, at this point, that the above statements
and what follows next only work for $2\le d\le4$, as is required by
the renormalizability  condition of the theory.

\section{Color symmetry breaking and gravity}

For the beginning of this section we forget the dynamical curved
space-time equivalent theory and turn back at the Euclidean theory
(\ref{pb1}).

\subsection{A background field}

Let us consider, again, the action (\ref{action4}). We suppose the
existence of a background field $\Upsilon$ which generates the
following field strength
\begin{equation}
F_{ijk}^{\phantom{ijk}l}(\Upsilon)=m^2(\delta_i^l\delta_{jk}-\delta_{ik}\delta_j^l)\;,\label{bg1}
\end{equation}
where $m$ has dimension of a mass. It is evident that the field
strength renormalizes according to the renormalization of the gauge
field and the coupling constant. Thus, a field strength for a
background would require that $m$ renormalizes nontrivially
according to the renormalization of the gauge field and the coupling
constant. In fact, according to Ap.\ref{ap3}, the renormalization
factor of $m^2$ is not independent and is given by
\begin{equation}
Z_{m^2}=Z_AZ_{\kappa}^2\;.\label{ren1}
\end{equation}\\\\ The condition (\ref{bg1}) implies that $\Upsilon$ is a solution
of the classical field equations if one requires the following extra
condition
\begin{equation}
\Upsilon_i^{\phantom{i}li}\delta_{jk}-\Upsilon_{ik}^{\phantom{ik}i}\delta^l_j+\Upsilon_{\phantom{l}kj}^{l}-
\Upsilon_{k\phantom{l}j}^{\phantom{k}l}=0\;.\label{bg2}
\end{equation}\\\\
It is not difficult to see that equation (\ref{bg1}) suggests that
$\Upsilon$ is a singular configuration. For that, we write
$\Upsilon$ as a pure gauge, which is just a gauge equivalent of the
vacuum configuration $\Upsilon'=0$,
\begin{equation}
\Upsilon_{ik}^{\phantom{ik}l}=\partial_i\theta_k^l\;.\label{pure1}
\end{equation}
And then, considering the linear approximation of equation
(\ref{bg1}), one finds
\begin{equation}
\left(\partial_i\partial_j-\partial_j\partial_i\right)\theta_k^l=m^2(\delta_i^l\delta_{jk}-\delta_{ik}\delta_j^l)\;.\label{topo01}
\end{equation}
This equation establishes the singular nature of $\Upsilon$ since
$\theta$ is clearly singular. An explicit example can be found in
\cite{Ribeiro:2003yn}, where $\Upsilon$ is associated with flux
tubes in an Abelian superconductor.

\subsection{Color symmetry breaking}

From (\ref{bg1}) we see that $\Upsilon$ explicitly breaks color
invariance since it mixes color and space-time indices. As a
consequence it will turn the color indices into observable
space-time indices. On the other hand, one might argue on the
physical consequences of the background in the sense that the field
strength, in the form (\ref{bg1}), is an exclusive solution of the
space-time representation. This means that this effect can be
described exclusively in the space-time representation. To answer
this question, we can take a look on the Noether current for global
color symmetry,
\begin{equation}
j_{pij}(A)=\frac{\delta
S_{YM}}{\delta(\partial_pA_{km}^{\phantom{km}n})}\delta_{(ij)}A_{km}^{\phantom{km}n}\;,\label{nother1}
\end{equation}
where, from (\ref{color2}),
\begin{equation}
\delta_{(ij)}A_{km}^{\phantom{km}n}=\delta_{mj}A_{ki}^{\phantom{ki}n}-\delta_{mi}A_{kj}^{\phantom{kj}n}\;.\label{var1}
\end{equation}
Thus
\begin{equation}
j_{pij}(A)=A_{ki}^{\phantom{ki}n}F_{pkj}^{\phantom{pkj}n}-A_{kj}^{\phantom{kj}n}F_{pki}^{\phantom{pki}n}\;.\label{nother2}
\end{equation}
For the background we have then
\begin{equation}
j_{pij}(\Upsilon)=m^2\left(\Upsilon_{ki}^{\phantom{ki}k}\delta_{jp}-\Upsilon_{kj}^{\phantom{kj}k}\delta_{ip}+
\Upsilon_{ijp}-\Upsilon_{jip}\right)\;.\label{nother3}
\end{equation}
Making use of condition (\ref{bg2}) we achieve, identically,
\begin{equation}
j_{pij}(\Upsilon)=0\;.\label{nother4}
\end{equation}
Thus, the background configuration lies at the nonphysical sector of
the theory. This means that the massless Goldstones states,
associated with the color symmetry breaking, should decouple from
the physical spectrum of the theory.\\\\ Obviously, considering the
gauge fixing term, (\ref{action3}), our conclusion remains the same
since the difference is just a BRST exact term
\begin{equation}
j_p^{\phantom{p}ij}(\Upsilon)=s[\mathcal{F}(\Upsilon,A,\bar{c},c,b)]\;,\label{brsexact}
\end{equation}
where $\mathcal{F}$ is a functional of the fields and the
background. In (\ref{brsexact}) the BRST variation of the background
is defined according to
\begin{equation}
s\Upsilon_k^a=f^{abc}\Upsilon_k^bc^c\;,
\end{equation}
or, in the space-time representation,
\begin{equation}
s\Upsilon_{ki}^{\phantom{ki}j}=\Upsilon_{ki}^{\phantom{ki}m}c^j_{\phantom{j}m}-
\Upsilon_{km}^{\phantom{km}j}c^m_{\phantom{m}i}\;.
\end{equation}
For the other fields the BRST transformations are given in
(\ref{brs1}).

\subsection{Effective action for the background}

Now, we write the gauge field as a perturbation around the
background
\begin{equation}
A_{ij}^{\phantom{ij}k}\rightarrow\Upsilon_{ij}^{\phantom{ij}k}+A_{ij}^{\phantom{ij}k}\;.\label{bg5}
\end{equation}
Thus,
\begin{equation}
F_{ijk}^{\phantom{ijk}l}(A)\rightarrow
F_{ijk}^{\phantom{ijk}l}(A)+m^2(\delta_{ik}\delta_j^l-\delta_i^l\delta_{jk})+\Upsilon_{i\phantom{m}k}^{\phantom{i}m}A_{jm}^{\phantom{jm}l}-
\Upsilon_{im}^{\phantom{im}l}A_{j\phantom{m}k}^{\phantom{j}m}-\Upsilon_{j\phantom{m}k}^{\phantom{j}m}A_{im}^{\phantom{im}l}+
\Upsilon_{jm}^{\phantom{jm}l}A_{i\phantom{m}k}^{\phantom{i}m}\;.\label{F1}
\end{equation}
To avoid terms depending on the background in (\ref{F1}) and,
consequently, in the resulting action, we write $\Upsilon$
as\footnote{Notice that the covariant derivative possess two color
indices. Thus, together with the ordinary space-time index, there
will be five indices in the space-time representation.}
\begin{equation}
\Upsilon_{ij}^{\phantom{ij}k}=\mathcal{F}_{ij}^{\phantom{ij}k}-D_{ijm}^{\phantom{ijm}kl}h_l^m\;,\label{tr1}
\end{equation}
in such a way that $\Upsilon$ is fixed in order to maintain the
relation (\ref{bg1}) while $\mathcal{F}$ and $h$ are arbitrary and
the covariant derivative is taken with respect the background
itself, $D=D(\Upsilon)$. Thus, we can get rid of the $\Upsilon$
dependent term in (\ref{F1}) by making the smooth limit of the
arbitrary functions $\mathcal{F}\leftrightarrow{D\cdot h}$ while
$F(\Upsilon)$ remains fixed and $\Upsilon$ becomes small. Thus, we
are taken the limit $\Upsilon\rightarrow0$ while keeping the
singular character of $\Upsilon$. This trick might be interpreted as
follows: A fixed background breaks the gauge invariance of the
background field method \cite{'t Hooft:1975vy}. However, in order to
control this breaking while keeping some gauge freedom on the
background, the fields $\mathcal{F}$ and $h$ were introduced in
(\ref{tr1}). As a consequence, the functions $\mathcal{F}$ and $h$
allow $\Upsilon$ to vary through a class of backgrounds that
generates (\ref{bg1}). Then, by making $\Upsilon$ as small as
possible, expression (\ref{F1}) reads
\begin{equation}
F_{ijk}^{\phantom{ijk}l}(A)\rightarrow
F_{ijk}^{\phantom{ijk}l}(A)+m^2(\delta_{ik}\delta_j^l-\delta_i^l\delta_{jk})\;.\label{F2}
\end{equation}
And the action (\ref{action4}) now reads
\begin{equation}
S_{\mathrm{eff}}=\frac{1}{4\kappa^2}\int{d^dx}\left[F_{ijk}^{\phantom{ijk}l}F^{ijk}_{\phantom{ijk}l}-4m^2F+2d(d-1)m^4\right]\;.\label{actiona8}
\end{equation}\\\\
We have introduced a new parameter, $m$, as a free parameter.
However, this parameter is not present in the starting action. Thus,
it might exist a condition to fix this parameter to a physical
consistent value. For instance, the mass parameter would be such
that it characterizes the vacuum nature of the background. To do so,
$m$ would be fixed, in a self consistent way, by requiring minimal
dependence in $m$ of the vacuum energy,
\begin{equation}
\frac{\partial\mathcal{W}}{\partial m^2}=0\;,\label{gap1}
\end{equation}
where the quantum action is defined as
\begin{equation}
e^{-\mathcal{W}}=\int{DADbD\bar{c}Dc}\;e^{-S_{\mathrm{eff}}-S_{gf}}\;.\label{func0}
\end{equation}
The gap equation (\ref{gap1}) in fact reads
\begin{equation}
\left<F\right>=d(d-1)m^2\;,\label{gap2}
\end{equation}
where $\left<F\right>$ is the expectation value of $F$ related to
the functional (\ref{func0}). The gap equation (\ref{gap2}) fixes
$m$ to a physical value $m_*$ which stabilizes the vacuum. We also
remark that this kind of gap equation (\ref{gap1}) have been
frequently used in Physics, in particular, in QCD where the gap
equation (\ref{gap1}) is imposed to find an optimal value for the
mass gap in QCD \cite{Sorella:2006ax}. Also in QCD, a similar gap
equation is used to fix the so called Gribov parameter, associated
with the improvement of the quantization of Yang-Mills theories
\cite{Gribov:1977wm,Zwanziger:1992qr,Dudal:2005na}.\\\\ Once $m_*$
is determined, the existence of the background turns out to be
completely characterized by $m_*$ and by the presence of the the
color breaking term $F$ in the action. Thus, after the computation
of $m^2_*$,
\begin{equation}
S_{\mathrm{eff}}=\frac{1}{4\kappa^2}\int{d^dx}\left[F_{ijk}^{\phantom{ijk}l}F^{ijk}_{\phantom{ijk}l}-4m_*^2F+2d(d-1)m_*^4\right]\;.
\label{action8}
\end{equation}
The theory is then described by the effective action
(\ref{action8}). We notice that in the action (\ref{action8}) the
parameters $\kappa$ and $m_*$ are assumed to be already fixed from
usual loop quantum computations \cite{Itzykson:1980rh}.\\\\
Notice that, due to color symmetry breaking mechanism, the group
indices are transformed into observable space-time indices. Thus
$A_{ij}^{\phantom{ij}k}$ would describe physical spin-1 excitations
as well as spin-2 physical excitations.

\subsection{Gravity}\label{grav}

Since we have now an effective theory, where a kind of
non-perturbative vacuum is consistently defined, we can perform the
mapping (\ref{map5}) and (\ref{map7}), providing now
\begin{equation}
S_{eff}^{\mathrm{e}}=\frac{1}{4\kappa^2}\int{d^dx}\;\mathrm{e}\left[R_{\mu\nu\alpha\beta}R^{\mu\nu\alpha\beta}-4m_*^2R+
2d(d-1)m_*^4\right]\;, \label{action9}
\end{equation}
where $R=R_{\mu\nu}^{\phantom{\mu\nu}\mu\nu}$.\\\\
The action (\ref{action9}) is $GL(d,\mathbb{R})$ invariant in the
space-time sector, and, due to the presence of the EH term, we can
associate it with gravity. In that sense, the mass parameter $m^2_*$
is related to both, Newton constant, $G$, and the cosmological
constant, $\Lambda$, through
\begin{eqnarray}
G&=&\frac{\kappa^2}{16\pi{m}_*^2}\;,\nonumber\\
\Lambda&=&\frac{d(d-1)}{2}m_*^2\;.\label{consts}
\end{eqnarray}
Thus,
\begin{equation}
S_{eff}^{\mathrm{e}}=\frac{1}{16\pi{G}}\int{d^dx}\;\mathrm{e}\left[\frac{d(d-1)}{8\Lambda}R_{\mu\nu\alpha\beta}R^{\mu\nu\alpha\beta}-R+
\Lambda\right]\;,\label{action9a}
\end{equation}
which is recognized as a generalization of the EH action in the
Palatini formalism.\\\\ Let us derive the field equations of action
(\ref{action9a}). For the connection $\Gamma$, the field equations
read
\begin{eqnarray}
&-&\mathcal{D}_\mu{R}^{\mu\nu\alpha}_{\phantom{\mu\nu\alpha}\beta}-2\left[\frac{d}{2}Q_\mu+T_{\mu\kappa}^{\phantom{\mu\nu}\kappa}\right]
R^{\mu\nu\alpha}_{\phantom{\mu\nu\alpha}\beta}+
T_{\mu\kappa}^{\phantom{\mu\kappa}\nu}R^{\mu\kappa\alpha}_{\phantom{\mu\kappa\alpha}\beta}\nonumber\\
&+&\frac{2\Lambda}{d(d-1)}\left[Q_\beta^{\phantom{\beta}\nu\alpha}-\delta^\nu_\beta\bar{Q}^\alpha+
2T^{\alpha\phantom{\beta}\nu}_{\phantom{\alpha}\beta}+U^{\mu\nu\alpha}_{\phantom{\mu\nu\alpha}\beta}\left(\frac{d}{4}Q_\mu+
T_{\mu\kappa}^{\phantom{\mu\kappa}\kappa}\right)\right]=0\;,\label{fe4}
\end{eqnarray}
where
\begin{equation}
U_{\mu\nu\alpha\beta}=g_{\mu\beta}g_{\nu\alpha}-g_{\mu\alpha}g_{\nu\beta}\;,\label{rbar}
\end{equation}
again, $Q_\mu$ is the Weyl convector (\ref{weyl1}), which can be
obtained from the non-metricity tensor $Q_{\mu\nu\alpha}$, defined
as
\begin{equation}
Q_{\mu\alpha\beta}=\mathcal{D}_\mu{g}_{\alpha\beta}\;.
\end{equation}
The quantity $\bar{Q}_\mu$ is also a vector constructed from the
non-metricity,
\begin{equation}
\bar{Q}_\mu=Q_{\nu\mu}^{\phantom{\mu\nu}\nu}\;.
\end{equation}
Finally, $T_{\mu\nu\alpha}$ is the torsion tensor (\ref{tor1}).\\\\
The variation with respect to the metric tensor provides
\begin{equation}
\frac{1}{2}g^{\mu\nu}R^{\sigma\rho\alpha\beta}R_{\sigma\rho\alpha\beta}-
R^\mu_{\phantom{\mu}\alpha\beta\gamma}R^{\nu\alpha\beta\gamma}-
R^{\phantom{\alpha\beta\gamma}\mu}_{\alpha\beta\gamma}R_{\phantom{\nu\alpha\beta\gamma}}^{\alpha\beta\gamma\nu}-
\frac{8\Lambda}{d(d-1)}\left[\frac{1}{2}g^{\mu\nu}(R-\Lambda)-R^{\mu\nu}\right]=0\;.\label{fe5}
\end{equation}
Contraction of this equation, (\ref{fe5}), with $g_{\mu\nu}$ results
on the trace equation
\begin{equation}
\frac{(d-4)}{4}R^{\sigma\rho\alpha\beta}R_{\sigma\rho\alpha\beta}+\frac{2\Lambda}{d(d-1)}\left(2-d\right)R+
\frac{2\Lambda^2}{(d-1)}=0\;.\label{fe6}
\end{equation}\\\\ Let us focus on the more interesting case $d=4$.
The field equations (\ref{fe4}) and (\ref{fe5}) now read
\begin{eqnarray}
-\mathcal{D}_\mu{R}^{\mu\nu\alpha}_{\phantom{\mu\nu\alpha}\beta}-2\left[2Q_\mu+T_{\mu\kappa}^{\phantom{\mu\nu}\kappa}\right]
R^{\mu\nu\alpha}_{\phantom{\mu\nu\alpha}\beta}+
T_{\mu\kappa}^{\phantom{\mu\kappa}\nu}R^{\mu\kappa\alpha}_{\phantom{\mu\kappa\alpha}\beta}& & \nonumber\\
+\frac{\Lambda}{6}\left[Q_\beta^{\phantom{\beta}\nu\alpha}-\delta^\nu_\beta\bar{Q}^\alpha+
2T^{\alpha\phantom{\beta}\nu}_{\phantom{\alpha}\beta}+U^{\mu\nu\alpha}_{\phantom{\mu\nu\alpha}\beta}\left(Q_\mu+
T_{\mu\kappa}^{\phantom{\mu\kappa}\kappa}\right)\right]&=&0\;,\nonumber\\
\frac{1}{2}g^{\mu\nu}R^{\sigma\rho\alpha\beta}R_{\sigma\rho\alpha\beta}-
R^\mu_{\phantom{\mu}\alpha\beta\gamma}R^{\nu\alpha\beta\gamma}-
R^{\phantom{\alpha\beta\gamma}\mu}_{\alpha\beta\gamma}R_{\phantom{\nu\alpha\beta\gamma}}^{\alpha\beta\gamma\nu}-
\frac{2\Lambda}{3}\left[\frac{1}{2}g^{\mu\nu}(R-\Lambda)-R^{\mu\nu}\right]&=&0\;,\nonumber\\
& &\label{fe9}
\end{eqnarray}
while the trace equation (\ref{fe6}) now provides
\begin{equation}
R=2\Lambda\;.\label{fe8}
\end{equation}
In the usual gravitation theory with cosmological constant,
\emph{i.e.}, the action (\ref{action9a}) with no quadratic terms in
the curvature, the equation (\ref{fe8}) is also obtained and the
solution is the $dS_4$ space-time where the curvature is given by
\begin{equation}
R_{\mu\nu\alpha\beta}=\frac{\Lambda}{6}U_{\mu\nu\alpha\beta}\;.\label{ds1}
\end{equation}
Remarkably, this expression is also an exact solution of
(\ref{fe9}). To see this one has to use the fact that the geometry
of the $dS_4$ space-time is Riemannian, $Q=T=0$.\\\\
It is worth mention that one would generate a gravity theory with
the $AdS_4$ as a solution if, in (\ref{gap1}), a negative value for
$m^2_*$ shows up. In that case $\Lambda<0$ and we are in fact facing
the $AdS_4$ instead of $dS_4$.

\section{Discussions and conclusions}

In this work we have established a mapping from an Euclidean $SO(d)$
gauge theory to a dynamical space-time with linear connection and
independent metric tensor. Moreover, in order to relate the mapping
with a model for gravity, we have developed a sketch for a color
symmetry breaking mechanism which generates spin-2 excitations in
Euclidean space-time. Thus, the mapping from the Euclidean space
with a gauge field to a curved space with linear connection provides
a modified gravity model with cosmological constant in the Palatini
formalism.\\\\ We noticed that that is exactly the mappings
(\ref{map5}) and (\ref{map7}) which leads action (\ref{action1})
into action (\ref{action6b}). However, we strongly remark again that
the presented color symmetry breaking mechanism is just a sketch.
Realistic quantum computations are needed in order to give estimates
for $\kappa$ and $m$ and, consequently, to $G$ and $\Lambda$. By
contrast, in principle, we have a vast freedom to find suitable
solutions for $\kappa$ and $m$ within the renormalization group
equations. This happens because, being renormalizable to all orders
in perturbation theory, the theory is scale invariant as well as
renormalization scheme invariant. Furthermore, the value the $m$ is
related to the value of $\Lambda$ which receives several
contributions from the non-perturbative vacuum energy of the rest of
fundamental interactions \cite{Shapiro:2006qx}. For instance, at the
electroweak sector there is a contribution coming from the Higgs
vacuum while from the QCD sector there would be a contribution from
the Gribov ambiguities and from condensates \cite{Dudal:2005na}.
Obviously, even if consistent solutions might not show up
analytically, numerical computations would be applied as well.
Another point is that the gap equation (\ref{gap1}) is an
\emph{ansatz} justified by the minimization principle of the vacuum
energy. It might be possible to exist another way to compute $m$ at
the quantum level. We emphasize that this kind of computation (for
$\kappa$ and $m$ within the renormalization group) is beyond the
scope of this article and must be analyzed \cite{work}. From now on,
for the rest of the conclusion we will assume that the computation
is
possible and gives reasonable values for $m$ and $\kappa$.\\\\
Concerning the space-time nature, we can conclude from the present
proposal that, for very high energies, the fundamental space-time
would be a continuous Euclidean space. Now, once we start to
decrease the energy, the background starts to show its presence,
breaking the color symmetry. Thus, spin-2 physical excitations show
up and the theory can be visualized as a dynamical curved
space-time. Under this scope, the fundamental space-time
configurations would be the $dS_4$ space (Or $AdS_4$ if
$\Lambda<0$).\\\\ Another point to be explored in the future would
be the inclusion of matter fields (non-gravitational fields) in the
model. In principle, we can include the standard model in this
formalism since we are dealing with an Euclidean space-time.
However, it should result, at the gravitational low energy regime,
in
a consistent quantum standard model in a classical curved space.\\\\
An evident interpretation of the method is that there is no
incompatibility between GR and QM. In fact, the quantum gravity
theory is described in an Euclidean space-time, thus, there are no
problems to be faced in the space and time definitions. At the
quantum level, the space and time are the usual quantum mechanical
parameters. At the classical level, the space and time are
translated into the GR space-time. However, quantum effects are no
longer associated with them. Thus, we can say that QG and GR would
be different sides of the same coin.\\\\ Finally, we recall
attention to the fact that the approach presented in this article
might be generalized to other groups but the $SO(d)$, as well as for
other classes of gauge theories. For instance, one may start with a
$GL(d,\mathbb{R})$ gauge theory and see if it can be casted into a
dynamical space-time. Obviously, in that particular case, the
space-time representation exists as well.

\section*{Acknowledgments}

The authors would like to acknowledge A.~Accioly, H.~J.~M.~Cuesta,
S.~A.~Dias, E.~L.~Rodrigues and J.~F.~Villas da Rocha for fruitful
and interesting discussions. Also J.~A.~Helay\"el-Neto is gratefully
acknowledge for fruitful and clarifying discussions, suggestions and
for reviewing the article. Still, we would like to kindly thank
T.~R.~B.~M.~Rodrigues for suggestions in the correction of the text.
Finally, the Conselho Nacional de Desenvolvimento Cient\'{i}fico e
Tecnol\'{o}gico (CNPq-Brazil) is acknowledged for financial support.

\appendix

\section{Renormalizability of the model}\label{ap1}

To study renormalizability of the action (\ref{action2}) we will use
the algebraic renormalization technique \cite{Piguet:1995er}. A
basic ingredient of this method relies on the so called BRST
symmetry. For the action (\ref{action2}) the BRST transformations
are
\begin{eqnarray}
sA_i&=&-D_i{c}\;,\nonumber\\
sc&=&\frac{1}{2}[c,c]\;,\nonumber\\
s\bar{c}&=&b\;,\nonumber\\
sb&=&0\;,\label{brs1}
\end{eqnarray}
where $s$, the BRST operator, is nilpotent $s^2=0$.\\\\ In order to
have a consistent quantum description of the model we introduce a
set of external BRST invariant sources coupled to the nonlinear BRST
variations,
\begin{eqnarray}
S_{ext}&=&s\mathrm{Tr}\int{d^dx}\left(-\Omega_i{A}_i+Lc\right)\nonumber\\
&=&\mathrm{Tr}\int{d^dx}\left(-\Omega_i{D}_i{c}+\frac{1}{2}L[c,c]\right)\;,\label{action1a}
\end{eqnarray}
where
\begin{equation}
s\Omega=sL=0\;.\label{brs2}
\end{equation}
Then, in this appendix we will work with the general BRST invariant
action
\begin{equation}
\Sigma=\frac{1}{4}\mathrm{Tr}\int{d^dx}F_{ij}F_{ij}+\mathrm{Tr}\int{d^dx}\;\left(b\;\partial_i{A}_i+
\bar{c}\;\partial_i{D}_i{c}\right)+\mathrm{Tr}\int{d^dx}\left(-\Omega_i{D}_i{c}+\frac{1}{2}L[c,c]\right)\;.\label{action2a}
\end{equation}\\\\
Let us take a look at the power counting renormalizability of
(\ref{action2}). For that, the UV dimensions of the fields and
sources are displayed in table \ref{table1a}. Observing the
dimension of the coupling constant we see that renormalizability
would be a feature of the restricted cases where $2\le d\le4$.
\begin{table}[t]
\centering
\begin{tabular}{|c|c|c|c|c|c|c|c|}
\hline
fields & $\partial$ & $s$ & $A$ & $c$ & $\bar{c}$ & $b$ & $\kappa$ \\
\hline
UV dimension & 1 & 0 & 1 & 0 & $d-2$ & $d-2$ & $(4-d)/2$ \\
Ghost number & 0 & 1 & 0 & 1 & $-1$ & 0 & 0\\
\hline
\end{tabular}
\caption{Quantum numbers of the relevant quantities.}
\label{table1a}
\end{table}

\subsection{Symmetries and Ward identities}

The symmetries of the model are described by the following Ward
identities:
\begin{itemize}

\item The Slavnov-Taylor identity
\begin{equation}
\mathcal{S}(\Sigma)=\mathrm{Tr}\int{d^dx}\left(\frac{\delta\Sigma}{\delta\Omega_i}\frac{\delta\Sigma}{\delta{A}_i}+
\frac{\delta\Sigma}{\delta{L}}\frac{\delta\Sigma}{\delta{c}}+b\frac{\delta\Sigma}{\delta\bar{c}}\right)=0\;.\label{st1}
\end{equation}

\item The ghost equation
\begin{equation}
\int{d^dx}\left(\frac{\delta\Sigma}{\delta{c}}+\left[\bar{c},\frac{\delta\Sigma}{\delta{b}}\right]\right)=
\int{d^dx}\left([\Omega_i,A_i]+[L,c]\right)\;.\label{ghost1}
\end{equation}

\item The gauge condition and the antighost equation
\begin{eqnarray}
\frac{\delta\Sigma}{\delta{b}}&=&\partial_i{A}_i\;,\nonumber\\
\frac{\delta\Sigma}{\delta\bar{c}}+\partial_i\frac{\delta\Sigma}{\delta\Omega_i}&=&0\;.\label{antighost1}
\end{eqnarray}

\item The $SL(2,\mathbb{R})$ symmetry
\begin{equation}
\mathrm{Tr}\int{d^dx}\left(c\frac{\delta\Sigma}{\delta\bar{c}}+\frac{\delta\Sigma}{\delta{L}}\frac{\delta\Sigma}{\delta{b}}\right)=0\;.\label{sl2r}
\end{equation}

\item Global color symmetry
\begin{equation}
\int{d^dx}\left(\left[A_i,\frac{\delta\Sigma}{\delta{A}_i}\right]+\left[c,\frac{\delta\Sigma}{\delta{c}}\right]+\left[\bar{c},
\frac{\delta\Sigma}{\delta\bar{c}}\right]+\left[b,\frac{\delta\Sigma}{\delta{b}}\right]+\left[\Omega_i,\frac{\delta\Sigma}{\delta\Omega_i}\right]+
\left[L,\frac{\delta\Sigma}{\delta{L}}\right]\right)=0\;.\label{color0}
\end{equation}

\end{itemize}

\subsection{Characterization of the most general counterterm}

Now, to show the renormalizability of the action (\ref{action2a}) we
will make use of the algebraic renormalization technique
\cite{Piguet:1995er}. For that, one adds to the classical action a
perturbation term,
\begin{equation}
\Sigma^{(1)}=\Sigma+\epsilon\Sigma^c\;,\label{quantumaction1}
\end{equation}
where the counterterm $\Sigma^c$ is an integrated polynomial on the
fields and sources with UV dimension up-bounded by four and
vanishing ghost number. The enforcement that the action
(\ref{quantumaction1}) obeys the Ward identities
(\ref{st1}-\ref{color0}) implies on the following constraints for
the counterterm,
\begin{eqnarray}
\mathcal{S}_\Sigma\Sigma^c&=&0\;,\nonumber\\
\int{d^dx}\frac{\delta\Sigma^c}{\delta{c}}&=&0\;,\nonumber\\
\frac{\delta\Sigma^c}{\delta{b}}&=&0\;,\nonumber\\
\frac{\delta\Sigma^c}{\delta\bar{c}}+\partial_i\frac{\delta\Sigma^c}{\delta\Omega_i}&=&0\;,\nonumber\\
\mathrm{Tr}\int{d^dx}\;c\frac{\delta\Sigma^c}{\delta\bar{c}}&=&0\;,\nonumber\\
\int{d^dx}\left(\left[A_i,\frac{\delta\Sigma^c}{\delta{A}_i}\right]+\left[c,\frac{\delta\Sigma^c}{\delta{c}}\right]+\left[\bar{c},
\frac{\delta\Sigma^c}{\delta\bar{c}}\right]+\left[\Omega_i,\frac{\delta\Sigma^c}{\delta\Omega_i}\right]+
\left[L,\frac{\delta\Sigma^c}{\delta{L}}\right]\right)&=&0\;.\label{constr1}
\end{eqnarray}
where $\mathcal{S}_\Sigma$ stands for the linearized Slavnov-Taylor
operator,
\begin{equation}
\mathcal{S}_\Sigma=\mathrm{Tr}\int{d^dx}\left(\frac{\delta\Sigma}{\delta\Omega_i}\frac{\delta}{\delta{A}_i}+
\frac{\delta\Sigma}{\delta{A}_i}\frac{\delta}{\delta\Omega_i}+
\frac{\delta\Sigma}{\delta{L}}\frac{\delta}{\delta{c}}+
\frac{\delta\Sigma}{\delta{c}}\frac{\delta}{\delta{L}}+b\frac{\delta}{\delta\bar{c}}\right)\;.\label{lst1}
\end{equation}
The first of (\ref{constr1}) defines a cohomological problem whose
general solution is given by
\begin{equation}
\Sigma^c=\Delta_0+\mathcal{S}_\Sigma\Delta^{-1}\;,\label{count1}
\end{equation}
where $\Delta_0$ is the nontrivial part of the cohomology,
\begin{equation}
\Delta_0\ne\mathcal{S}_\Sigma\cdot(\mathrm{something})\;,
\end{equation}
and $\mathcal{S}_\Sigma\Delta^{-1}$ is the trivial part. The
nontrivial part is an integrated polynomial on the fields and
sources with UV dimension up-bounded by four and vanishing ghost
number while $\Delta^{-1}$ is an integrated polynomial with UV
dimension up-bounded by four and ghost number given by $-1$. In
fact, it is straightforward to show that the most general
counterterm (\ref{count1}) is determined by \cite{Piguet:1995er}
\begin{equation}
\Delta_0=a_0S_{\mathrm{YM}}\;,
\end{equation}
and
\begin{equation}
\Delta^{-1}=a_1\mathrm{Tr}\int{d^dx}\left(\Omega_i+\partial_i\bar{c}\right)A_i\;.
\end{equation}
where $a_0$ and $a_1$ are independent renormalization parameters.

\subsection{Quantum stability}

The last step to prove renormalizability is to show that the
counterterm (\ref{count1}) can be reabsorbed in the classical action
(\ref{action2a}) by means of multiplicative redefinition of the
fields, sources and parameters according to
\begin{equation}
\Sigma(\phi_0,J_0,\kappa_0)=\Sigma(\phi,J,\kappa)+\epsilon\Sigma^c(\phi,J,\kappa)\;,\label{abs1}
\end{equation}
where
\begin{eqnarray}
\phi_0&=&Z^{1/2}_\phi\phi\;,\nonumber\\
J_0&=&Z_JJ\;,\nonumber\\
\kappa_0&=&Z_\kappa\kappa\;,\label{abs2}
\end{eqnarray}
and
\begin{eqnarray}
\phi&\in&\{A,b,c,\bar{c}\}\;,\nonumber\\
J&\in&\{\Omega,L\}\;.\label{abs3}
\end{eqnarray}
In fact it is not difficult to show that the action (\ref{action2a})
is multiplicativelly renormalizable, where the independent
renormalization factors are given by
\begin{eqnarray}
Z_\kappa&=&1-\epsilon\frac{a_0}{2}\;,\nonumber\\
Z_A^{1/2}&=&1+\epsilon\left(\frac{a_0}{2}+a_1\right)\;,
\end{eqnarray}
and the nonindependent renormalization factors read
\begin{eqnarray}
Z_c^{1/2}=Z_{\bar{c}}^{1/2}=Z_\Omega^{-1/2}&=&Z_\kappa^{-1/2}Z_A^{-1/4}\;,\nonumber\\
Z_b^{1/2}=Z_L^{-2}&=&Z_A^{-1}\;.
\end{eqnarray}\\\\ This ends the proof of the renormalizability of pure $SO(d)$ gauge
theory in a $2\le d\le4$-dimensional Euclidean space-time.

\section{Renormalization of $m^2_*$}\label{ap3}

In this appendix we provide the program for derivation of the
renormalization rule of the background parameter $m^2_*$. We omit
the details since they are somewhat simple to be performed. For that
we consider as a starting point the fact that the background is a
particular configuration of the gauge field $A$. Thus
\begin{equation}
\Upsilon_0=Z_A^{1/2}\Upsilon\;.
\end{equation}
Now, perturbing the equation (\ref{abs1}) according to (\ref{bg5})
we will find an expression where the mass parameter $m^2_*$ is
present. Thus, performing the usual multiplicative renormalization
(\ref{abs2}-\ref{abs3}) together with
\begin{equation}
{m^2_*}_0=Z_{m^2}m^2_*\;,
\end{equation}
we find expression (\ref{ren1}).

\end{document}